\documentclass[aps,pra,letterpaper,twocolumn,floatfix,longbibliography]{revtex4-2} 
\usepackage{amsmath,amssymb} 
\usepackage[T1]{fontenc}
\usepackage[utf8]{inputenc} 

\usepackage{graphicx} 
\usepackage{float}

\usepackage{hyperref}
\hypersetup{colorlinks=true,citecolor={black},linkcolor={blue},urlcolor={blue}} 
\usepackage{xcolor}
\usepackage{bm}
\usepackage{bbold}

\newcommand{\delone}{\frac{\partial}{{\partial}r}\left(\frac{\partial}{{\partial}r}+\frac{1}{r}\right)}
\newcommand{\pdr}{\partial}

\newcommand{\ve}{\varepsilon}

\graphicspath{{.}{./figs/}}

\begin{document}

\title{Spin-orbital effect on polariton state in traps}

\author{Yuri G. Rubo} 
\email{ygr@ier.unam.mx}
\affiliation{Instituto de Energías Renovables, Universidad Nacional Autónoma de México, Temixco, Morelos, 62580, Mexico} 

\begin{abstract}
I discuss similitude and differences of spin-orbital effects for electrons in quantum wells with the Rashba coupling 
and for polaritons in semiconductor microcavities with TE-TM splitting. 
Contrary to the case of electron, the ground state of polariton in the trap can be non-degenerate and can possess specific polarization structure. 
For the case of azimuthally symmetric trap and sufficiently strong spin-orbital coupling, the ground state is either radial or azimuthal vortex, depending on the sign of the coupling constant. The effect is strongly enhanced for polaritons trapped in a ring, where even weak TE-TM splitting results in formation of vorticity and definite polarization of the ground state. 	
The Hamiltonian for quasi-1D motion of polaritons in the ring is derived and it is shown the the dispersion of polaritons depend qualitatively on the curvature of the ring.  
\end{abstract}

\date{\today}

\maketitle

	\section{Introduction}\label{sec:intro}
The Rashba spin-orbital coupling (SOC) \cite{rashba59,rashba60} and the discovery of electric-dipole spin resonances \cite{rashba61,rashba90} played a pivotal role in our understanding of spin-dependent phenomena in bulk semiconductors. The role of SOC is even more pronounced in low dimensional semiconductor structures, where the Rashba \cite{vasko79,bychkov84} and the Dresselhaus \cite{dresselhaus55} terms, which are both linear in the wave vector in quantum wells, allow effective manipulation of electron spins by electric field \cite{rashba03}. These discoveries have founded in part the spintronics \cite{zutic04,awschalom13lib}, have huge influence on developing of quantum computing with semiconductor quantum dots, and have pronounced impact on other areas of condensed matter physics \cite{manchon15}.

This influence resulted especially evident and straightforward in the development of research and applications of exciton-polariton condensates in semiconductor microcavities. Strong coupling of light with excitons in semiconductors was first discussed in pioneering work by S.~I.~Pekar in 1957 \cite{Pekar57}, which has led him to discovery of additional light waves, and quanta of these waves are referred to as polaritons nowadays. Polaritons, due to their excitonic component, are interacting quasiparticles, both with phonons and themselves, and they are sensitive to external fields, which in principle allow their manipulation in the crystal. While efficient relaxation of polaritons in bulk crystals can take place, the possibility of Bose-Einstein condensation in its original sense is not possible: The lower polariton branch is photon-like at low energies and it does not have the ground state to condense to \cite{note1}. (The same argument is known for the blackbody radiation \cite{huang87,klaers10}.) The situation is different in two-dimensional (2D) semiconductor microcavities \cite{kavokin17lib}, where polariton condensates have been discovered \cite{kasprzak06,balili07,baumberg08} and intensively investigated in recent years. 

Semiconductor microcavities are made of a few-microns-thick semiconductor layer placed between two distributed Bragg mirrors to hold a specific mode of light trapped in the growth direction. Several semiconductor quantum wells are also put in the antinodes of the light mode, in order to provide strong coupling of the light with excitons in quantum wells, resulting in formation of lower and upper polariton branches. As long as only the lower polariton branch is of interest, the polariton states in a microcavity are similar to the 2D electron states, in sense that both are described by two-component wave functions. 
Many polarization effects and phenomena observed and discussed for the polariton system are, on the one hand, to some extent copycat of those for electrons in quantum wells, but, on the other hand, they frequently bring in new ideas related to higher space and time coherence of polaritons and to feasibility experimental observation and verification by optical means. It is worth mentioning the optical spin Hall effect \cite{kavokin05,leyder07}, polariton Berry-phase interferometer \cite{shelykh09}, and the polariton \emph{zitterbewegung} \cite{sedov18,sedov21} 
as some examples of effects that rely heavily on the presence of strong coupling between orbital and pseudospin dynamics in polariton transport (see also review \cite{shelykh10} and references therein for additional information). 
The linear in wave vector SOC can be regarded as a vector potential with non-commuting components, or the non-Abelian gauge field, and the related effects can be clearly demonstrated by optical means \cite{whittaker21}. 

More recently, there have been growing interest in applications of polariton condensates for analog and quantum computation. 
The task of forming well defined two-level system with polaritons is noticeably more complex than it is for electrons in semiconductor low-dimensional quantum systems. Even in single-particle approximation the energy levels of a trapped polariton are qualitatively different from those of an electron. The main difference comes from the role played by the time-reversal symmetry (TRS). For electrons the time-reversal operator $\mathcal{T}_e$ is odd, $\mathcal{T}_e^2=-1$, which ensures at least double degeneracy of all electron levels in the case when TRS is present. This fact, known as the Kramers theorem \cite{sachs87lib}, allows, in principle, to address a particular Kramers doublet for qubit operations. In contrast, the polaritons are composite bosons and the time-reversal operator for them $\mathcal{T}_p$ is even, $\mathcal{T}_p^2=1$, which does not guarantee the degeneracy of energy levels. 

The condition of TRS imposes limitations on possible spin-orbit synthetic Hamiltonians \cite{rechcinska19}, which can be obtained for polaritons, and limit substantially the analogy between electron and polariton systems. This is clear reflection of the fact that polaritons are described by pseudospin, rather than the real spin. In particular, the Rashba term cannot appear in the polariton Hamiltonian with preserved TRS. Interestingly, however, the so-called Rashba-Dresselhaus term, which corresponds to the case when the Rashba and the Dresselhaus contributions have the same amplitude \cite{bernevig06}, is allowed for both electrons and polaritons, which provides identical Hamiltonians with SU(2) symmetry and allows formation of persistent spin helix \cite{bernevig06,krol21}. Having this in mind, the term ``spin-orbital coupling'' is used for polariton system in what follows. 

In this paper, the energy levels of polaritons in azimuthally symmetric traps are considered. It is shown that the ground state can become a singlet in the case when the SOC is strong enough. The ground-state singlet is characterized by well-defined radial or azimuthal linear polarization. 
This effect turns out to be especially strong for polaritons trapped in the rings, and since this configuration has received much interest recently \cite{manni11,dreismann14,liu15,lukoshkin18,mukherjee21}, the SOC for the motion of polaritons in the ring is studied in more details.
To highlight the effects of SOC, the consideration is carried out neglecting dissipation, i.e., assuming infinite life-time of polaritons in the traps. This approximation is acceptable for ultra-high-quality microcavities \cite{mukherjee21,estrecho21}, with the polariton lifetimes exceeding $100\,\mathrm{ps}$.

The equations for the case of SOC due to transverse-electric-transverse-magnetic (TE-TM) splitting in axially symmetric traps are formulated in Sec.\ \ref{sec:basic} below. The energy levels of parabolic and hard-wall disk traps are considered in Sec.\ \ref{sec:dots}. Sec.\ \ref{sec:ring} is devoted to the important case of ring traps. The conclusions are given in Sec.\ \ref{sec:concl}.

\section{Basic equations}\label{sec:basic}
A polariton mode in microcavity is fully described by the in-plane component of the electric field $\mathbf{E}$, and it is convenient to use the 2D complex vector $\bm{\psi}=\{\psi_x,\psi_y\}\propto\mathbf{E}$ normalizing it to the number of polaritons belonging to the mode. In the cylindrically symmetric case, the kinetic energy density near the bottom of the lower polariton branch should contain two invariants, $|\bm{\nabla}\cdot\bm{\psi}|^2$ and $|\bm{\nabla}\times\bm{\psi}|^2$. 
As a result, the Schrödinger equation is written as 
\begin{multline}\label{SchEqVec}
	-\frac{\hbar^2}{2m_t}(\bm{\nabla}\cdot\bm{\nabla})\bm{\psi}
	-\frac{\hbar^2}{2}\left(\frac{1}{m_l}-\frac{1}{m_t}\right)\bm{\nabla}(\bm{\nabla}\cdot\bm{\psi}) \\
	=[E-U(\mathbf{r})]\bm{\psi}.
\end{multline}
Here $U(\mathbf{r})$, with $\mathbf{r}=\{x,y\}$ being the 2D position, is the potential energy and $m_{l,t}$ are the longitudinal and the transverse effective masses of polaritons, respectively. For free polaritons, the transverse mass $m_t$ corresponds to the transverse-electric (TE) mode with $(\bm{\nabla}\cdot\bm{\psi})=0$, while the longitudinal mass $m_l$ corresponds to the transverse-magnetic (TM) mode with $[\bm{\nabla}\times\bm{\psi}]=0$.

Instead of vector $\bm{\psi}$ they usually use two circular polarization components of the field $\psi_{\pm1}$ defined by 
\begin{equation}\label{PsiCircComp}
	\bm{\psi}=\frac{\hat{\mathbf{x}}+i\hat{\mathbf{y}}}{\sqrt{2}}\psi_{+1}
	         +\frac{\hat{\mathbf{x}}-i\hat{\mathbf{y}}}{\sqrt{2}}\psi_{-1},
\end{equation}
and combine them into the column $\Psi=(\psi_{+1},\psi_{-1})^\mathsf{T}$. The $2\times2$ Hamiltonian corresponding to Eq.\ \eqref{SchEqVec} is then
\begin{equation}\label{MatrHam}
	H=\frac{\hbar^2}{2m^*}\begin{pmatrix} k_-k_+ & {\gamma}k_-^2 \\[0.5em] {\gamma}k_+^2 & k_+k_- \end{pmatrix} 
	+ U(\mathbf{r})\mathbb{1},
\end{equation}
where $\mathbf{k}=-i\bm{\nabla}$, $k_\pm=k_x{\pm}ik_y$, and
\begin{equation}\label{MassGamma}
	\frac{1}{m^*}=\frac{m_t+m_l}{2m_tm_l}, \qquad \gamma=\frac{m_t-m_l}{m_t+m_l}.
\end{equation}  
One can see that the difference of transverse and longitudinal masses sets SOC of polariton modes. This is the basic SOC in the system and it is essentially the same as for the pure light (see \cite{bliokh15} for a review).
The splitting parameter $\Delta_\mathrm{LT}=\hbar\gamma/m^*$ is also used to define the strength of SOC in stead of dimensionless coupling constant $\gamma$. 

In the case of azimuthally symmetric potential energy $U(r)$, that is mainly considered below, the wave functions can be written in the general form 
\begin{equation}\label{WaveFun}
	\Psi=\begin{pmatrix}e^{i(m-1)\phi}\,f(r) \\[0.5em] e^{i(m+1)\phi}\,g(r)\end{pmatrix},
\end{equation}
where the integer $m=0,\pm1,\pm2,\dots$ is the winding number and $\phi$ is the azimuth angle. Using the relations
\begin{equation}\label{KpmRule}
	k_{\pm}e^{im\phi}f(r)=-ie^{i(m\pm1)\phi}\left(\frac{d}{dr}\mp\frac{m}{r}\right)f(r)
\end{equation}
one obtains 
\begin{widetext}
\begin{subequations}\label{EqsRadial}
\begin{equation}
	-\frac{\hbar^2}{2m^*}\left(\frac{d}{dr}+\frac{m}{r}\right)\left(\frac{d}{dr}-\frac{m-1}{r}\right)f(r)
	-\frac{\gamma\hbar^2}{2m^*}\left(\frac{d}{dr}+\frac{m}{r}\right)\left(\frac{d}{dr}+\frac{m+1}{r}\right)g(r)+U(r)f(r)=Ef(r),
\end{equation}
\begin{equation}
	-\frac{\gamma\hbar^2}{2m^*}\left(\frac{d}{dr}-\frac{m}{r}\right)\left(\frac{d}{dr}-\frac{m-1}{r}\right)f(r)
	-\frac{\hbar^2}{2m^*}\left(\frac{d}{dr}-\frac{m}{r}\right)\left(\frac{d}{dr}+\frac{m+1}{r}\right)g(r)+U(r)g(r)=Eg(r).
\end{equation}	
\end{subequations}
\end{widetext}
for the radial functions $f(r)$ and $g(r)$. 

Independently of particular form of $U(r)$ some conclusions about degeneracy of the energy levels can be drawn on the basis of time-reversal symmetry. Note that the time-reversal for the linear polarization components is reduced to complex conjugation, as it should be for two independent oscillators, $\mathcal{T}_p\{\psi_x,\psi_y\}=\{\psi_x^*,\psi_y^*\}$. 
Therefore, in circular polariton basis one has $\mathcal{T}_p(\psi_{+1},\psi_{-1})^\mathsf{T}=(\psi_{-1}^*,\psi_{+1}^*)^\mathsf{T}$ or $\mathcal{T}_p=K\sigma_x$, using the Pauli matrices $\sigma_{x,y,z}$ and the complex-conjugation operation $K$. This is in strike contrast to the electron case, where $\mathcal{T}_e=K\sigma_y$. 

Applying $\mathcal{T}_p$ to the wave functions \eqref{WaveFun} and taking into account that the solutions of \eqref{EqsRadial} can be chosen to be real functions one has 
\begin{equation}\label{TRSRule}
	m\,{\rightarrow}-m, \qquad f\,{\rightarrow}\,g, \qquad g\,{\rightarrow}\,f.
\end{equation}
These substitutions leave the system (\ref{EqsRadial}a,b) unchanged and this imply that the energies of the states $m$ and $-m$ for $m\ne0$ coincide. The $m=0$ states, however, are the eigenstates of $\mathcal{T}_p$ and corresponding energy levels remain non-degenerate.   

\begin{figure}[t]
	\centering 
	\includegraphics[width=0.48\textwidth]{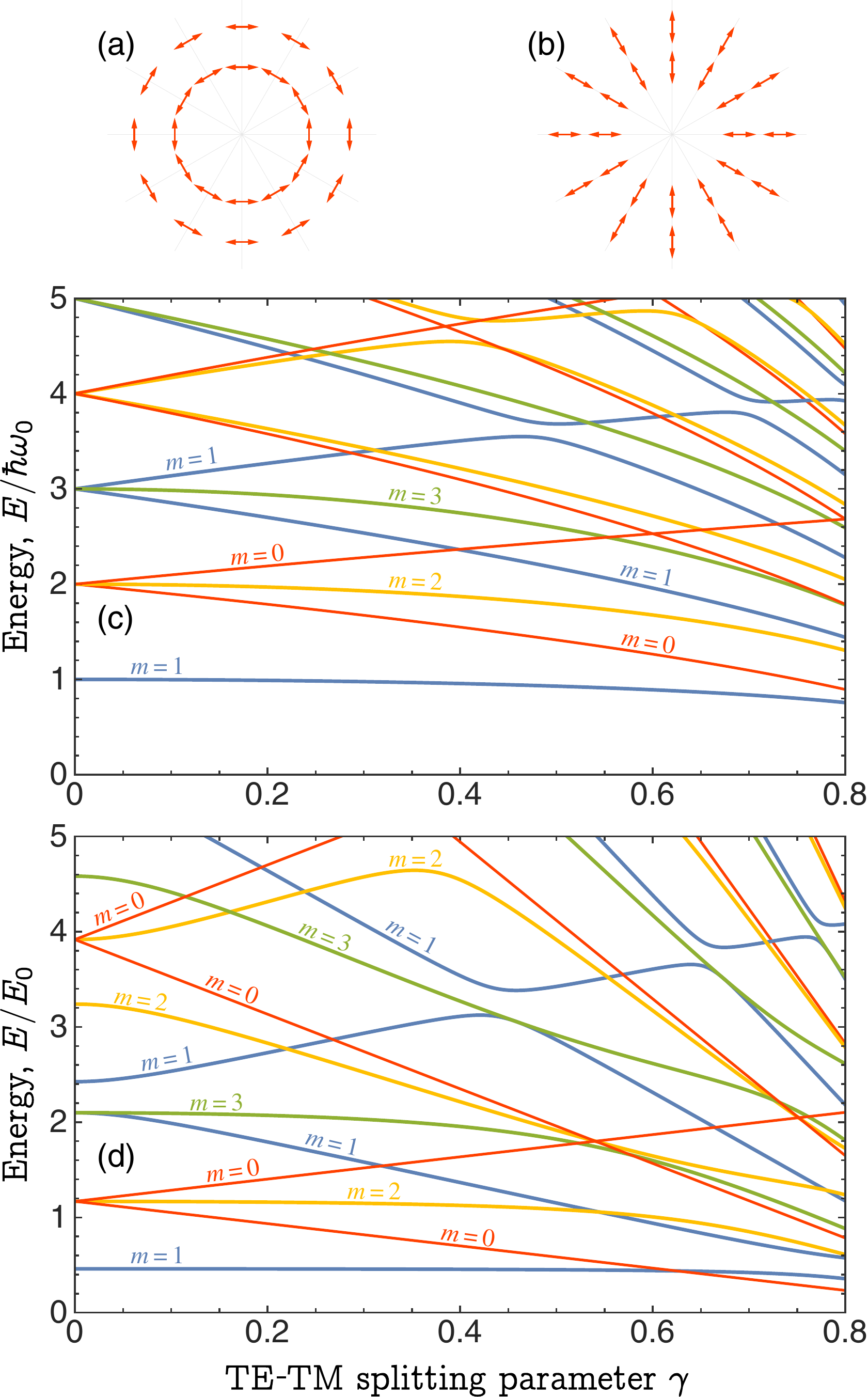} 
	\caption{
	Showing the linear polarization of the states with $m=0$, where the winding number $m$ 
	is defined by Eq.\ \eqref{WaveFun}. The polarization is either azimuthal (a) or radial (b).   
	(c) Energy levels for a 2D harmonic trap with the frequency $\omega_0$. 
	(d) The energy levels for a hard-wall disk with the radius $R_0$, given in the units of $E_0=\hbar^2/2{\pi}R_0^2m^*$.
	The schematic geometry of microcavity is shown in the inset.
	Thick lines show the double-degenerate levels with $|m|>0$, and thin (red) lines show the energies of linearly polarized 
	non-degenerate states (a,b) with $m=0$. 
	}
    \label{fig:1}
\end{figure}

The singlet solutions for $m=0$ possess well-defined polarization patterns of polariton field. In this case we have $g(r)={\pm}f(r)$, so that polarization is linear, and the radial function satisfies the equation for single-component system with angular momentum equal to one,
\begin{equation}\label{RadialMono}
	-\frac{\hbar^2}{2m_{t,l}}\frac{d}{dr}\left(\frac{d}{dr}+\frac{1}{r}\right)f(r)+U(r)f(r)=Ef(r).
\end{equation}
Here the solutions with $g=-f$ are characterized by the transverse mass $m_t=m^*/(1-\gamma)$ and by the azimuthal polarization field shown in Fig.\ \ref{fig:1}(a), while the solutions with $g=f$ possess the longitudinal mass $m_l=m^*/(1+\gamma)$ and the radial polarization field shown in Fig.\ \ref{fig:1}(b).

The presence of these pure longitudinal and pure transverse states in the traps plays an important role in propagation of polariton wave packets and leads to possibility of polarization rectification \cite{sedov19}. Note also that in general case of arbitrary potential of the dot $U(\mathbf{r})$, e.g., for random polariton billiards, the singlet polariton states are described by a real valued vector field $\bm{\psi}(\mathbf{r})$ and they are thus linearly polarized everywhere in the trap.

\section{Polariton levels in traps}\label{sec:dots}
A quantum trap for polaritons can be approximated with a parabolic potential near its bottom, and it is natural to begin with description of energy levels for the case of harmonic potential $U(r)=m^*\omega_0^2r^2/2$. The dependence of energies on the TE-TM splitting parameter $\gamma$, that measures the extent of longitudinal and transverse mass difference  \eqref{MassGamma}, is shown in Fig.\ \ref{fig:1}(c). 
The presence of SOC lifts the degeneracies of 2D harmonic oscillator levels and the mean distance between the levels decreases with increasing $\gamma$. The overall behavior of energy levels resembles the case of parabolic quantum dots with the Rashba SOC \cite{rashba12}. In our case the decrease of mean level spacing is because the increase of $|\gamma|$ with fixed $m^*$ corresponds to the increase of the longitudinal effective mass $m_l$ for $\gamma<0$ or to the increase of the longitudinal mass $m_t$ for $\gamma>0$. In what follows we consider the latter case only. 
Note that on the one hand the change $\gamma{\rightarrow}-\gamma$ corresponds to the exchange of $m_l$ and $m_t$. On the other hand, $\gamma{\rightarrow}-\gamma$ can be compensated by the change $g{\rightarrow}-g$ in the system \eqref{EqsRadial}. This implies 90 degree rotation of polarization plane, so that the azimuthal vortices in Fig.\ \ref{fig:1}(a) transform to the radial ones in Fig.\ \ref{fig:1}(b) and vice versa. 

While the energies of the twofold degenerate states with $|m|>0$ are defined by some effective masses residing in between of $m_{l,t}$, the energies of non-degenerate radial vortices are set by the lightest mass $m_l$ and the energies of the azimuthal vortices are defined by the heaviest mass $m_t$. The energies of the former states grow rapidly with increasing TE-TM parameter $\gamma$, while the energies of the latter states decrease and approach the energy of the $m=1$ state, which is the ground state for $\gamma=0$. As a result, the ground state changes at some critical value of SOC parameter $\gamma_c$ and the lowest-energy azimuthal vortex becomes the ground state for $\gamma>\gamma_c$. In the case of parabolic potential this happens at very high value of $\gamma_c\simeq0.918$ [not shown in Fig.\ \ref{fig:1}(c)].

The effect of ground states change is more pronounced in the case polariton confinement in hard-wall disk. For the case of Rashba SOC this problem has been solved in Refs.\ \cite{bulgakov01,tsitsishvili04,rashba12}. The hard-wall billiards in general and an axially symmetric disk trap in particular can be formed by excitation of polaritons along the perimeter by a spatially structured pump \cite{askitopoulos13,cristofolini13,gao15}, so that the polaritons move freely inside the trap and are reflected from the boundary due to repulsion from the incoherent excitonic reservoir.  
The energy levels of polaritons in the disk are shown in Fig.\ \ref{fig:1}(d). In this case the crossing of levels takes place at $\gamma_c\simeq0.625$. This value is still too high as compared to typical values of $\gamma\sim0.1$ in semiconductor microcavities \cite{kavokin04}. Note, however, the values of $\gamma$ can be substantially bigger in microcavities with organic layers \cite{stelitano09}. The effect of the ground state change can be observed for small values of TE-TM splitting in the case of trapping polaritons in the ring, which is considered in the next Section.

\begin{figure}[t]
	\centering 
	\includegraphics[width=0.48\textwidth]{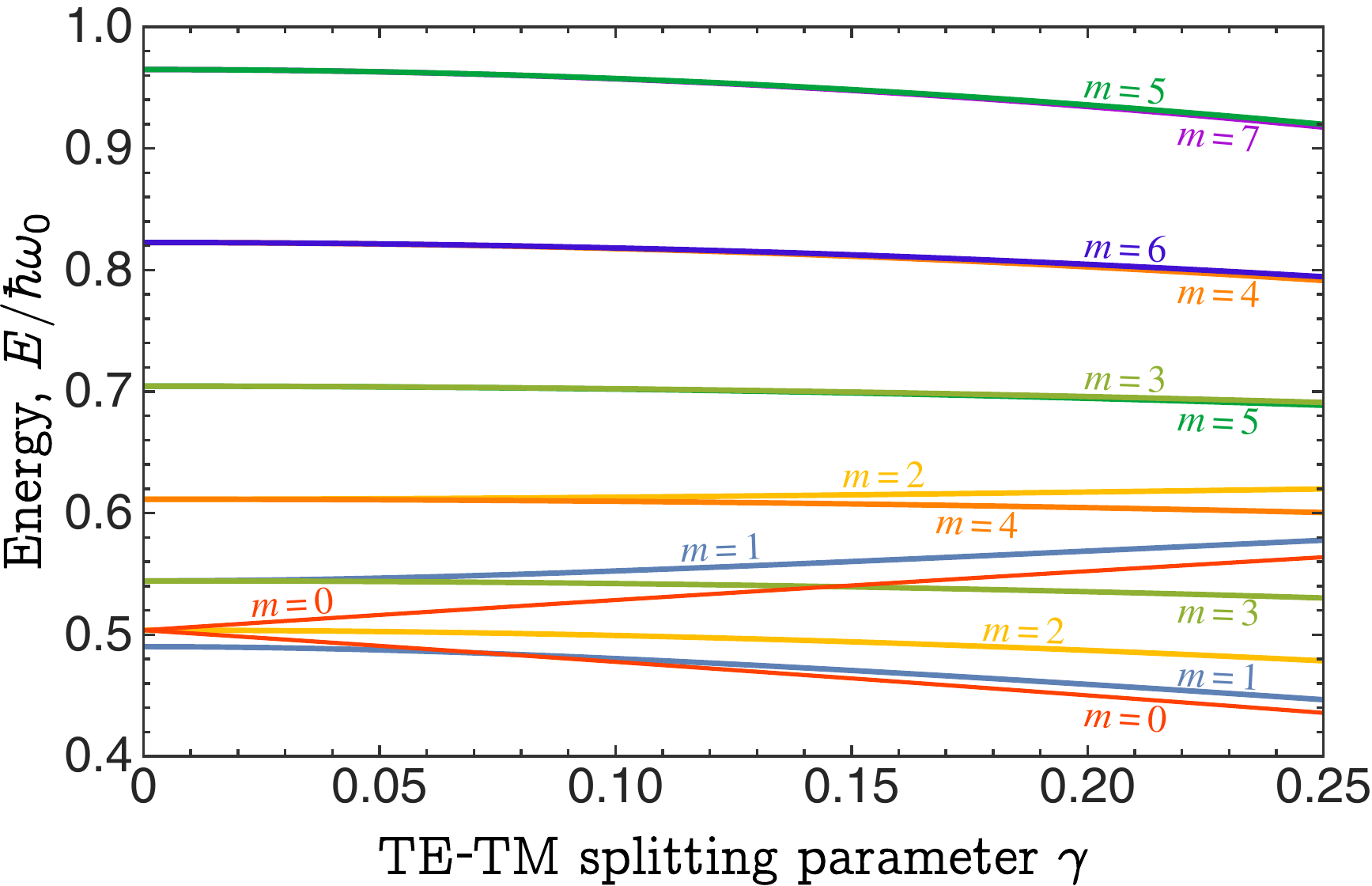} 
	\caption{   
	Energy levels for polariton moving in the ring (see the beginning of Sec.\ \ref{sec:ring} for details). 
	The schematic geometry of microcavity is shown in the inset.
	The confining frequency in radial direction $\omega_0=40\hbar/m^*a^2$, where $a$ is the radius of the ring. 
	}
    \label{fig:2}
\end{figure}

\section{Polaritons in the ring}\label{sec:ring}
The energy levels for polaritons moving in the ring are presented in Fig.\ \ref{fig:2}. The calculations have been carried our for a simple model potential $U(r)=U_0[1-2(r/a)^2+(r/a)^4]$, where $a$ is the radius of the ring and $U_0$ defines the barrier in the center. Since quasi-1D motion is achieved in the limit of large $U_0\gg\hbar^2/m^*a^2$, the levels shown in Fig.\ \ref{fig:2} are practically identical to the levels in harmonic approximation for the potential with
\begin{equation}\label{HarmPotRing}
	U(r)=\frac{1}{2}m^*\omega_0^2(r-a)^2,
\end{equation} 
where $\omega_0=\sqrt{8U_0/m^*}$. This frequency is used to scale the energy levels in Fig.\ \ref{fig:2}. One can observe that the crossing of low-energy levels takes place at a rather small value of TE-TM splitting parameter, $\gamma_c\simeq0.074$ for $\omega_0m^*a^2/\hbar=40$. Moreover, the crossing value $\gamma_c$ decreases further with increase of ring confinement frequency. The effect of TE-TM splitting on the high-energy levels is much lees pronounced. The levels remain nearly double-degenerate for large winding numbers $m$. Due the the form of the wave function defined by Eqs.\ \eqref{PsiCircComp} and \eqref{WaveFun}, where the angular momenta of the $\psi_{\pm1}$ components differ by 2, the separations of the first (lowest energy) level with $m=6$ and the second level with $m=4$, as well as of the second $m=5$ level and the first $m=7$ level, are nearly invisible.

To get better understanding of the effect of SOC on the low-energy levels it is convenient to derive the effective Hamiltonian for the quasi-1D angular motion of a polariton in the ring. This Hamiltonian can be particularly useful for the analysis of polariton condensation in the ring geometry, spontaneous azimuthal currents \cite{nalitov17}, formations of space-time periodic polarization patterns \cite{nalitov19,sedov21,mukherjee21}, and annular Josephson vortices \cite{munozmateo20,chestnov21}. The annular Hamiltonian can be obtained in adiabatic approximation by averaging over appropriate radial wave function. In the presence of TE-TM splitting of polariton bands this procedure should be done carefully, since there are two masses, $m_t$ and $m_l$, and it is not evident which one should be used for calculation of the radial component of the wave function. (Obtaining of the annular Hamiltonian for the case of Rashba SOC is also error-prone, see discussion in Ref.\ \cite{meijer02}.)

The derivation can be conveniently carried out using the equations for the radial $\psi_r$ and azimuthal $\psi_\phi$ components of the vector function $\bm{\psi}$, 
\begin{subequations}\label{PolarComp}
\begin{align}
	\psi_r &=\frac{1}{\sqrt{2}}\left(e^{i\phi}\psi_{+1}+e^{-i\phi}\psi_{-1}\right), \\
	\psi_\phi &=\frac{i}{\sqrt{2}}\left(e^{i\phi}\psi_{+1}-e^{-i\phi}\psi_{-1}\right).
\end{align}
\end{subequations}
From Eq.\ \eqref{SchEqVec} one obtains
\begin{widetext}
\begin{subequations}\label{EqsPolar}
\begin{equation}\label{EqsPolara}
	-\frac{\hbar^2}{2}\left[\frac{1}{m_l}\delone+\frac{1}{m_tr^2}\frac{\pdr^2}{\pdr\phi^2}\right]\psi_r	
	+\frac{\hbar^2}{m^*}\left[\frac{1}{r^2}\frac{\pdr}{\pdr\phi}-\frac{\gamma}{r}\frac{\pdr^2}{{\pdr}r\pdr\phi}\right]\psi_\phi
	=\left[E-U(r,\phi)\right]\psi_r, 
\end{equation}
\begin{equation}\label{EqsPolarb}
	-\frac{\hbar^2}{m^*}\left[\frac{1}{r^2}\frac{\pdr}{\pdr\phi}+\frac{\gamma}{r}\frac{\pdr^2}{{\pdr}r\pdr\phi}\right]\psi_r
	-\frac{\hbar^2}{2}\left[\frac{1}{m_t}\delone+\frac{1}{m_lr^2}\frac{\pdr^2}{\pdr\phi^2}\right]\psi_\phi
	=\left[E-U(r,\phi)\right]\psi_\phi.
\end{equation}
\end{subequations}
\end{widetext}
Application of adiabatic approximation implies separation of fast motion in radial direction and slow annular motion, $\psi_r=R_r(r)\Phi_r(\phi)$ and $\psi_\phi=R_\phi(r)\Phi_\phi(\phi)$, and it is clear from Eqs.\ \eqref{EqsPolar} that the radial part $R_r$ is defined by Eq.\ \eqref{RadialMono} with the longitudinal mass $m_l$, while the azimuthal part $R_\phi$ is given by the same equation, but with the transverse mass $m_t$. For parabolic trapping potential \eqref{HarmPotRing}, averaging over radial components can be carried out using scaling relation between $R_r$ and $R_\phi$. 
The resulting equations for the angular components are
\begin{subequations}\label{EqsAnnul}
\begin{align}
	\left[E_l(\gamma)+\frac{\hbar^2k_\phi^2}{2m_t}\right]\Phi_r+i\varkappa(\gamma)\frac{\hbar^2k_\phi}{m^*a}\Phi_\phi &=E\Phi_r, \\
	-i\varkappa(\gamma)\frac{\hbar^2k_\phi}{m^*a}\Phi_r+\left[E_t(\gamma)+\frac{\hbar^2k_\phi^2}{2m_l}\right]\Phi_\phi &=E\Phi_\phi.
\end{align}
\end{subequations}
Here $k_\phi=-i\partial/a\partial\phi$ is the annular wave vector, 
$E_t(\gamma)=\ve_0\sqrt{1-\gamma}$ and $E_l(\gamma)=\ve_0\sqrt{1+\gamma}$ with $\ve_0\simeq(\hbar\omega_0/2)+(\hbar^2/2m^*a^2)$ are the energies of vortex solutions shown in Figs.\ \ref{fig:1}(a) and \ref{fig:1}(b), respectively, and they are plotted in Fig.\ \ref{fig:2} as $m=0$ solutions. Finally, the coefficient 
\begin{equation}
	\varkappa(\gamma)=\sqrt{\frac{2(1-\gamma^2)^{1/4}}{(1+\gamma)^{1/2}+(1-\gamma)^{1/2}}}
\end{equation} 
appears due to the overlap of the radial functions $R_r(r)$ and $R_\phi(r)$. If the difference of longitudinal and transverse masses is not too extreme, this coefficient is very close to 1 and it can be safely omitted from Eqs. \eqref{EqsAnnul}. 

\begin{figure}[t]
	\centering 
	\includegraphics[width=0.48\textwidth]{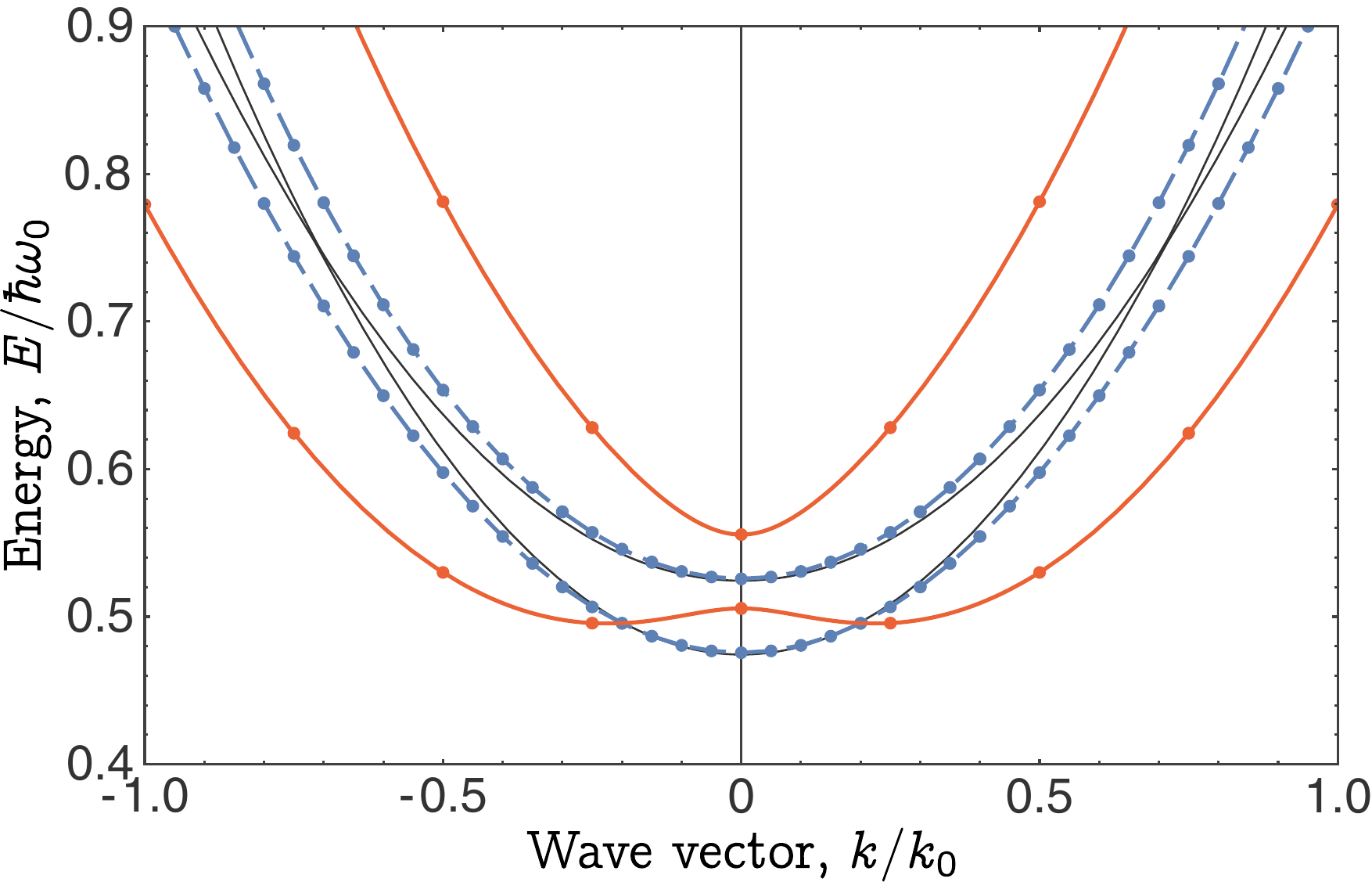} 
	\caption{   
	Showing the dispersion of polariton bands in the wire for the TE-TM splitting parameter $\gamma=0.1$ and 
	different values of the wire radius $a$.
	Quantized values of the wave vector, which is given in the units of $k_0=\sqrt{m^*\omega_0/\hbar}$, are shown by dots. 
	Thick solid and dashed lines corresponds to $k_0a=4$ and $k_0a=20$, respectively. 
	Thin solid lines show the dispersion in the straight wire ($a\rightarrow\infty$). 
	}
    \label{fig:3}
\end{figure}

The energy levels $E_m$ can be found by replacing $k_\phi\rightarrow m/a$ and Eq.\ \eqref{EqsAnnul} describes 
very well these levels in the adiabatic limit $\hbar^2m^2/2m^*a^2\ll\ve_0$. For small $|\gamma|\ll1$ one has
\begin{equation}\label{EqEm}
	E_m=\ve_0+\frac{\hbar^2m^2}{2m^*a^2}
	\pm\sqrt{\left(\frac{\gamma\ve_0}{2}\right)^2+\left(\frac{\hbar^2m}{m^*a^2}\right)^2}\,.
\end{equation} 

It is evident that even weak TE-TM splitting cannot be neglected when considering the motion of polaritons in the ring, since the small parameter $\gamma$ is multiplied by a large energy $\ve_0$ that scales quantization of levels in radial direction. The point of level crossing corresponds to very small $\gamma_c\ll1$, and it can be obtained from \eqref{EqEm} that
\begin{equation}\label{GammaC}
	\gamma_c=\frac{3\hbar}{m^*a^2\omega_0}.
\end{equation} 

Another way to view these results relies on the properties of polariton dispersion in quantum wire. The $2\times2$ Hamiltonian defined by Eq.\ \eqref{EqsAnnul} implies that in the straight wire, i.e., for $a\rightarrow\infty$, there are two independent modes. Polariton can possess transverse polarization, in which case it is quantized in the wire with longitudinal mass and moves along the wire with the transverse mass, or vice versa. For a curved wire, however, these two modes are coupled. The coupling is linear in the wave vector, similarly to the Rashba coupling. Remarkably, the coupling strength for polaritons is proportional to the curvature of the wire $a^{-1}$ and can be therefore controlled. \cite{[{Note that weak coupling between these modes, also linear in wave vector, can appear in a straight but asymmetric wire, see\ }]shelykh18} 

The polariton bands in the wire are shown in Fig.\ \ref{fig:3}. One can see that there is qualitative change in the lowest band dispersion with increasing radius of the wire, related to disappearance of two side minima. This corresponds to the change of double degenerate ground state to the non-degenerate one. Note also that, contrary to the electron case, the spin-orbital coupling produces splitting of the bands at $k=0$. 

The polarization vortex is expected to be the ground state for the polariton rings investigated of Ref.\ \cite{mukherjee19}, with the ring radius of $a=50\,\mu\mathrm{m}$ and the localization length in the radial direction corresponding to $\sqrt{\hbar/m\omega_0}\simeq4\,\mu\mathrm{m}$. In this case, the level crossing value $\gamma_c\simeq0.019$. For the TE-TM splitting parameter $\gamma=0.1$, the distance to the first excited ($m=1$) level is about $0.2\,\mu\mathrm{eV}$, which is about an order of magnitude smaller than the dissipation broadening of the levels at the polariton lifetime $\tau\sim200\,\mathrm{ps}$ \cite{mukherjee21}. 
To improve level resolution it is necessary to increase localization in the radial direction. Efficient spin-flip relaxation is also necessary for spontaneous formation of polarization vortices shown in Fig.\ \ref{fig:1}(a,b). However, the polarization vortices can be directly excited with corresponding Laguerre-Gaussian beams. Depending on the sign of $\gamma$, one of them will be stable, both topologically and energetically, while the other should decay into the lower energy polariton states. In general, at least three polariton states, the $m=0$ singlet and the $m=1$ doublet, should be taken into account considering polariton condensation in the presence of dissipation and spin-dependent polariton-polariton interaction.

\section{Conclusions}\label{sec:concl}
Spin-orbital coupling (TE-TM splitting) for exciton-polaritons has pronounced effect on their levels in traps and can change the symmetry of the ground state. The effect is especially strong for polaritons localized in the ring traps, where the effective 1D Hamiltonian of their motion possesses linear in wave vector coupling between two polarization modes, similarly to the case of the Rashba SOC. In this case even weak SOC cannot be neglected. 

The crossover from the double-degenerate ground state in the ring to the singlet polarization-vortex state occurs at rather small values of SOC coupling strength, that is inversely proportional the confining frequency of the 1D wire and to the square of the ring radius, see Eq.\ \eqref{GammaC}. As a result, the polaritons form the polarization-vortex singlet state in large-radius rings, while the ground state in small-sized rings is a doublet without particular polarization structure. 

Importance of TE-TM splitting for the polariton motion in curved wires discussed in this paper is only one little example from the rich world of different physical phenomena based upon existence of linear in wave vector spin-orbital coupling in solids. The seed, that gave rise to this magnificent tree of new ideas and effects, was planted by Emmanuel Rashba more than 60 years ago. It is spectacular how, after all these years, this tree continues to grow and bear new fruits. 

\section*{Acknowledgments}\label{sec:acknow}
This work was supported in part by CONACYT (Mexico) Grant No.\ 251808 and by PAPIIT-UNAM Grant No.\ IN106320.


%
%
%

\end{document}